\documentclass[aps, reprint,superscriptaddress,groupedaddress]{revtex4-1}  

\usepackage{hyperref} 
\hypersetup
{
    colorlinks=true,       
    linkcolor=blue,        
    citecolor=blue,        
    urlcolor=blue           
} 

\usepackage{amsmath} 
\usepackage{amssymb}   
\usepackage{bm}
\usepackage{graphicx} 
\usepackage{dcolumn} 
 
\usepackage{subfigure} 

\let\revappendix\appendix

\usepackage[capitalize]{cleveref}

\begin{document}
\title{Intrinsic spin-orbit torque in an antiferromagnet with a weakly noncollinear spin configuration}

\author{Suik Cheon} 
\author{Hyun-Woo Lee}
\email{hwl@postech.ac.kr}
\affiliation{Department of Physics, Pohang University of Science and Technology, Pohang 37673, Korea}

\date{\today}

\begin{abstract}
An antiferromagnet is a promising material for spin-orbit torque generation. Earlier studies of the spin-orbit torque in an antiferromagnet are limited to collinear spin configurations. We calculate the spin-orbit torque in an antiferromagnet whose spin ordering is weakly noncollinear. Such noncollinearity may be induced spontaneously during the magnetization dynamics even when the equilibrium spin configuration is perfectly collinear. It is shown that deviation from perfect collinearity can modify properties of the spin-orbit torque since noncollinearity generates extra Berry phase contributions to the spin-orbit torque, which are forbidden for collinear spin configurations. In sufficiently clean antiferromagnets, this modification can be significant. We estimate this effect to be of relevance for fast antiferromagnetic domain wall motion.
\end{abstract}

\maketitle

\let\oldhat\hat
\renewcommand{\hat}[1]{\oldhat{\mathbf{#1}}} 
\renewcommand{\vec}[1]{\mathbf{#1}}


\section{Introduction}

An antiferromagnet (AFM) has been used in spintronic devices to generate the exchange bias.
Recently there are ongoing efforts to utilize other properties of AFM; AFMs can exhibit much faster magnetization dynamics (in THz scale) than ferromagnets (FMs) do (in GHz scale) and AFM devices do not suffer from the mutual interference unlike FM counterparts \cite{reviewAFM}.

Of particular interest in this respect is the AFM's ability to generate the spin-orbit torque (SOT).
SOT is an electrically generated torque through the spin-orbit coupling (SOC) 
and its study has been limited mostly to FM bilayers \cite{dmsSOT} made of a nonmagnetic heavy metal layer and a FM metal layer \cite{PhysRevB.72.033203,*PhysRevB.77.214429,*PhysRevB.78.212405}.
Thus theoretical \cite{PhysRevLett.113.157201,PhysRevB.95.014403} and experimental \cite{Wadleyaab1031} confirmation of AFM's ability to generate SOT opened an alternative path towards electrically controlled spintronic devices.
SOT generated by AFM CuMnAs has been used to electrically switch the N\'{e}el order of the AFM itself \cite{Wadleyaab1031}.
It has been suggested \cite{PhysRevLett.117.017202,*PhysRevLett.117.087203} that electrically controlled AFM devices may exhibit more superb functionalities than FM counterparts do.
To materialize AFM-based device applications, it is desired to understand properties of SOT in AFM clearly.
In this respect, further efforts to clarify these properties are necessary both theoretically and experimentally.

\begin{figure}[t!]
     \subfigure{\label{fig:LAFM}}
     \subfigure{\label{fig:BAFM}}
  \includegraphics[width=8.5cm]{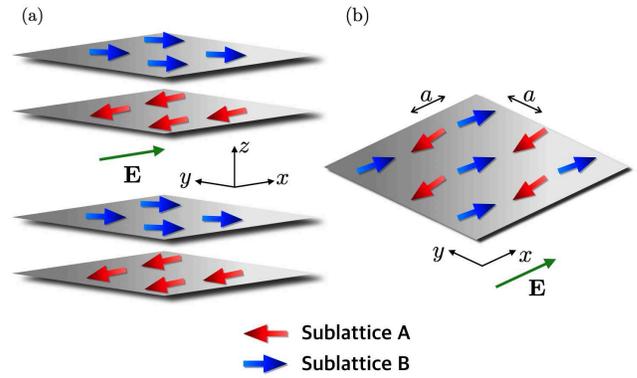}
\caption{Schematics of the (a) layered AFM structure and the (b) bipartite AFM structure. When spin configurations are collinear, these systems have antiunitary symmetries (see text).}
\label{fig:AFMstructure}
\end{figure}

In this paper, we examine theoretically effects of the noncollinearity on SOT in AFM. 
This study is motivated by the observation that even in AFMs with collinear spin configurations in equilibrium, their spin configurations become noncollinear and develop a weakly ferromagnetic component during the magnetization dynamics \cite{PhysRevB.83.054428}. 
On the other hand, existing theoretical studies \cite{PhysRevLett.113.157201,PhysRevB.95.014403} of SOT in AFM deal only with collinear spin configurations, probably because the dynamically generated noncollinearity is usually small. 
We investigate how the noncollinearity affects SOT and find that in the ``weak'' ferromagnetic regime, where the ferromagnetic component is sufficiently smaller than a threshold, the ferromagnetic component affects SOT only minor way. 
But in the ``strong'' ferromagnetic regime, where the ferromagnetic component is sufficiently larger than a threshold, the ferromagnetic component does modify SOT significantly. 
The threshold value depends on the level broadening of energy levels and temperatures.
Interestingly, it is found that for weakly disordered AFMs with weak level broadening, the threshold value at room temperatures for the crossover from the ``weak'' ferromagnetic regime to the ``strong'' ferromagnetic regime can be very low and the ``strong'' ferromagnetic regime can be realized even when the ferromagnetic component is much smaller than the antiferromagnetic component, implying the sensitivity of SOT to the ferromagnetic component and the noncollinearity. 
This is the main finding of this paper. 
We also demonstrate that this sensitivity originates from the symmetry breaking caused by the noncollinearity.
As exemplified in various examples, symmetries often impose constraints on the way Berry phase effects emerge. 
A recent example is the symmetry-breaking-induced anomalous behavior of the anomalous Hall effect in noncollinear AFMs \cite{PhysRevLett.112.017205,*Kubler:2014eb,*Kiyohara:2015jx,*Nayak:2016fr,*Suzuki:2016ft}.
When spin configurations are collinear, AFMs often have antiunitary symmetries whereas the symmetries are broken once spin configurations become noncollinear (Fig. \ref{fig:AFMstructure}).
We show that when spin configurations are collinear, the antiunitary symmetries prevent certain types of interband Berry-phase processes from contributing to SOT whereas when spin configurations are noncollinear, such contraint from the antiunitary symmetries disappears and previously ``forbidden'' interband Berry-phase processes generate extra contributions to SOT.
In disordered AFMs, where the level broadening $\Sigma$ by disorder is larger than the energy correction by the noncollinearity, the extra contributions are small.
But in clean AFMs, where $\Sigma$ is smaller than the energy correction by the noncollinearity, the extra contributions can be sizable.

The paper is organized as follows: 
In Sections \ref{subsec:LAFM} and \ref{subsec:BAFM}, we introduce the layered and bipartite AFM models, and those symmetry properties. 
We demonstrate that the symmetry of AFMs with collinear spin configuration forbids particular channels of Berry phase contributions to the SOT.
Next, we show that the symmetry breaking due to noncollinear spin configurations generates additional SOTs, which are fobidden for collinear spin configurations.
The detailed symmetry analysis for the bipartite AFM is given in the Appendix. 
In Sec. \ref{sec:Discussion}, we discuss the results, and the paper is summarized in \cref{sec:Summary}.


\section{\label{sec:Result}Result}

For demonstration of additional Berry phase effect due to noncollinearity, we examine two types of AFMs shown in Fig. \ref{fig:AFMstructure}.
The layered AFM structure in Fig. \ref{fig:LAFM} remains invariant under the time-reversal followed by the space inversion if its spin configuration is collinear.
This combined operation of the two is an antiunitary symmetry of the system.
The bipartite AFM structure in Fig. \ref{fig:BAFM} also has an antiunitary symmetry operation, which consists of the time-reversal followed by the translation of distance $a$ along $x$ or $y$ direction.
This symmetry also holds only when its spin configuration is collinear.
In both types of AFMs, the antiunitary symmetries are broken when spin configurations become noncollinear.

\subsection{\label{subsec:LAFM}Layered AFM}

\begin{figure}[t!]  
\includegraphics[width=8.5cm]{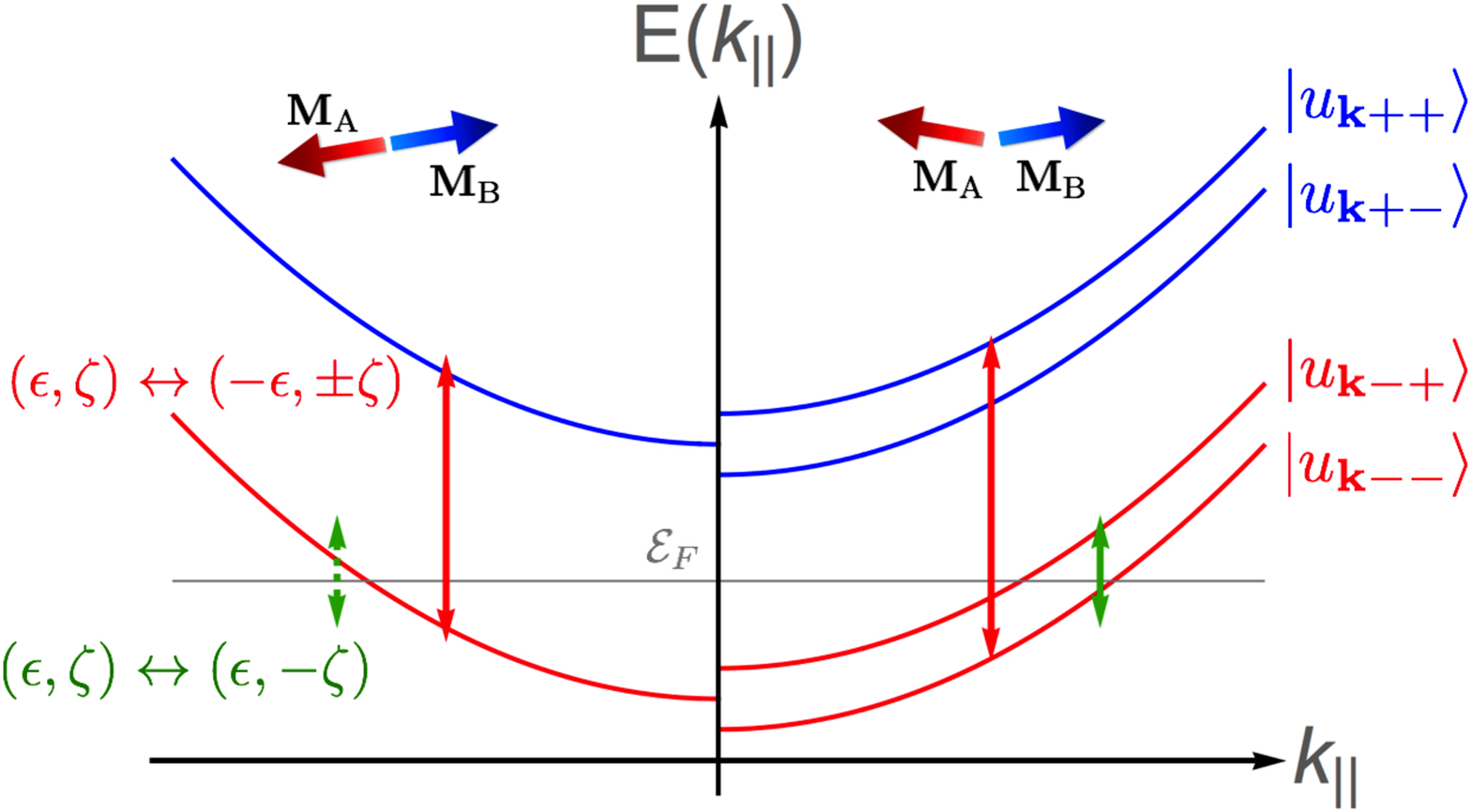}  
\caption{
The schematic energy dispersion relation for layered AFM. Here the vertical axis is energy, and the horizontal axis is momentum along $x$, or $y$ direction. $\mathcal{E}_F$ denotes Fermi energy. In the left panel (collinear case, $\vec{M}_{\text{A}} = - \vec{M}_{\text{B}}$), each state is degenerated due to the Kramers theorem. In the right panel (noncollinear case, $\vec{M}_{\text{A}} \neq - \vec{M}_{\text{B}}$), however, the degeneracy is lifted by the FM order exchange term. Each arrow indicates the interband transition in Eq. \eqref{inter}. Here, the green arrows indicate the interband process, which are prohibited by the PT symmetry for $\vec{M}_{\text{A}} = - \vec{M}_{\text{B}}$ but allowed when the PT symmetry is broken for $\vec{M}_{\text{A}} \neq - \vec{M}_{\text{B}}$.
}
\label{fig:band}
\end{figure}

We investigate layered AFM (Fig. \ref{fig:LAFM}) first.
Materials such as Mn$_2$Au \cite{doi:10.1063/1.3003878,*Barthem:2013go} belong to this structure.
Schematically speaking, the layered AFM consists of two sublattices A and B (Fig. \ref{fig:LAFM}) that form alternating layers stacked along the $z$-direction.
Note that the sublattices A and B are subject to the inversion symmetry breaking of opposite sign.
In the presence of the sublattice magnetizations $\vec{M}_{\text{A}}$ and $\vec{M}_{\text{B}}$ with $| \vec{M}_{\text{A}} | = | \vec{M}_{\text{B}} | = 1$, the system does not remain invariant under the inversion P nor under the time-reversal T.
But the system does remain invariant under the combined operation PT if $\vec{M}_{\text{A}} = - \vec{M}_{\text{B}}$.
This symmetry is antiunitary since $($PT$)^2 = -1$, and induces two-fold degeneracy to the energy dispersion.
On the other hand, if $ \vec{M}_{\text{A}}$ and $\vec{M}_{\text{B}}$ are not exactly collinear and FM order $\vec{m} \equiv (\vec{M}_{\text{A}} + \vec{M}_{\text{B}})/2 $ is not zero, PT is not a symmetry and the degeneracy is lifted (Fig. \ref{fig:band}).
Such degeneracy lifting can affect the Berry phase contribution to SOT \cite{kurebayashi2014antidamping,*PhysRevB.91.144401} significantly as we demonstrate below.

In order to study SOT, we examine the following four-band Hamiltonian for the layered AFM structure \cite{PhysRevB.93.180408,*PhysRevB.94.054409}.
%
\begin{equation}\label{Eq:modelHcolAFM}
 \mathcal{H}_{\text{L}} 
  = \frac{ \hbar^2 \vec{k}^2_{\parallel} }{2 m_e^*} \text{I}_\tau +
 \begin{pmatrix}
     J \bm{\sigma} \cdot \vec{M}_{\text{A}} & \gamma(k_z) \\
     \gamma^* (k_z) & J \bm{\sigma} \cdot \vec{M}_{\text{B}} 
 \end{pmatrix}_\tau 
 + \alpha \bm{\sigma} \cdot ( \vec{k} \times \hat{z} ) \tau_z,
\end{equation}
%
where $m_e^*$ is the effective electron mass within the $xy$ plane, $\vec{k}_{\parallel} = ( k_x , k_y , 0)$ is the in-plane component of the Bloch momentum $\vec{k}$, $\text{I}_\tau$ is a $2\times 2$ identity matrix in the sublattice space, $\gamma(k_z)$ is a weak hopping energy along $z$ direction, $J$ is an exchange interaction parameter, $\alpha$ is Rashba SOC parameter, $\tau_z$ is $z$ component of Pauli matrix which is $\pm 1$ for the sublattice A/B, and $\bm{\sigma}$ is the Pauli matrix for electron spin.
Note that the Rashba SOC term contains $\tau_z$ since A and B sublattices are subject to the opposite signs of the inversion symmetry breaking.
In terms of the N\`{e}el order $\vec{n} \equiv (\vec{M}_{\text{A}} - \vec{M}_{\text{B}} )/2$ and the FM order $\vec{m}$, the second term of $\mathcal{H}_{\text{L}}$ may be expressed as $J \bm{\sigma} \cdot \vec{n} \tau_z + J \bm{\sigma} \cdot \vec{m} \text{I}_\tau + \text{Re}[ \gamma(k_z)] \tau_x - \text{Im}[ \gamma(k_z) ] \tau_y $.
Under PT, $\vec{k} \to  \vec{k}$, $\bm{\sigma} \to - \bm{\sigma}$, $\tau_z \to - \tau_z$, $\tau_x \to \tau_x$, $\text{Im}[ \gamma(k_z) ] \tau_y \to  \text{Im}[ \gamma(k_z) ] \tau_y$, I$_\tau \to $I$_\tau$.
Thus $J \bm{\sigma} \cdot \vec{m} \text{I}_\tau$ is the only term in $\mathcal{H}_{\text{L}}$ that changes sign under PT and breaks the PT symmetry.

Calculation of the energy eigenvalues is straightforward.
For each $\vec{k}$, there are four energy eigenvalues, which we label by $\epsilon, \zeta = \pm$.
For $\vec{m} = 0$, $E_{\vec{k} \epsilon \zeta}(\vec{m} = 0) =  \frac{ \hbar^2 \vec{k}_\parallel^2 }{2 m_e} + \epsilon \sqrt{ \Delta^2_\vec{k}  + \gamma^2(k_z) }$.
%
Note that $E_{\vec{k} \epsilon \zeta} $ is independent of $\zeta$ and thus doubly degenerate.
Here $\Delta_\vec{k} = | J \vec{n} + \alpha \vec{k} \times \hat{z} |$.
For $\vec{m} \neq 0$, the degeneracy is lifted.
When the FM order exchange energy scale $J | \vec{m}|$ is smaller than the Rashba SOC energy scale $\alpha k_F$, where $k_F$ is the Fermi wave vector, the degeneracy-lifted eigenvalues read
%
\begin{eqnarray}\label{Eq:EvalueFM}
 &&E_{\vec{k} \epsilon \zeta}( \vec{m})  =  \frac{ \hbar^2 \vec{k}^2_{\parallel} }{2 m_e} + \epsilon  \sqrt{ \Delta^2_\vec{k}  + \gamma^2(k_z) } \nonumber  \\
  &&
  \phantom{asdfaasdsdasdsdfas} 
  +  \zeta   J  | \vec{m}| \xi_{\vec{k}}^{\text{L}} 
 + \mathcal{O} \bigl (|\vec{m}|^2 \bigl),
\end{eqnarray}
%
where $(\xi_{\vec{k}}^{\text{L}})^2 =  (  \gamma^2(k_z) \phantom{.} + \phantom{.} \alpha^2 |  \hat{m} \cdot ( \vec{k} \times \hat{z})|^2)/ (  \Delta^2_\vec{k}  + \gamma^2(k_z) ) $.
Note that the level spacing between $ E_{\vec{k}, \epsilon,  + }( \vec{m})$ and $ E_{\vec{k}, \epsilon,  - }( \vec{m})$ is proportional to not only $\vec{m}$ but also $\xi_{\vec{k}}^{\text{L}}$.
The energy dispersion relations for $\vec{m}=0$ and $\vec{m} \neq 0$ are depicted schematically in Fig. \ref{fig:band}.

When a constant electric field $\vec{E}$ is applied to the system, there appear non-equilibrium spin densities $\delta \vec{S}_{\text{A}}$ and $\delta \vec{S}_{\text{B}}$ at the two sublattices A and B, which in turn generates SOT $\vec{T}_\varsigma = ( J/\hbar ) \vec{M}_{\varsigma} \times \delta \vec{S}_{\varsigma} $, where $\varsigma = $ A or B.
$\delta \vec{S}_{\varsigma}$ contains two contributions, $\delta \vec{S}_{\varsigma} = \delta \vec{S}_{\varsigma}^{\text{Berry}} + \delta \vec{S}_{\varsigma}^{\text{non-Berry}}$, where $\delta  \vec{S}_{\varsigma}^{\text{Berry}}$ denotes the contribution arising from the Berry phase and $\delta \vec{S}_{\varsigma}^{\text{non-Berry}}$ denotes extrinsic contributions \cite{PhysRevLett.113.157201,PhysRevB.95.014403,kurebayashi2014antidamping,*PhysRevB.91.144401, PhysRevB.91.134402}, which are not related to the Berry phase but instead rely on scattering processes.
For instance, the Rashba Edelstein effect and the spin relaxation-induced nonadiabatic processes can contribute to $\delta \vec{S}^{\text{non-Berry}}_{\zeta}$.
Here we focus on $\delta  \vec{S}_{\varsigma}^{\text{Berry}}$ since the other contribution $\delta  \vec{S}_{\varsigma}^{\text{non-Berry}}$ is less sensitive to $\vec{m}$ and thus $\vec{m}$ may be regarded as zero for small $\vec{m}$ as far as $\delta  \vec{S}_{\varsigma}^{\text{non-Berry}}$ is concerned.
On the other hand, $\delta  \vec{S}_{\varsigma}^{\text{Berry}}$ is more sensitive to $\vec{m}$ as we demonstrate now.
According to the Kubo formula, $\delta  \vec{S}_{\varsigma}^{\text{Berry}}$ may be expressed as 
%
\begin{eqnarray} 
	&& \delta   \vec{S}^{\text{Berry}}_{\varsigma}
	 =
	  \frac{  \hbar}{V} \sum_{\vec{k}, \epsilon  \zeta  \neq \epsilon' \zeta'}
	   \text{Re}\biggl[   \frac{   f_{\vec{k} \epsilon  \zeta } - f_{\vec{k} \epsilon'  \zeta' }   }{(  E_{\vec{k} \epsilon  \zeta } - E_{\vec{k}\epsilon'  \zeta' } + i \Sigma)^2} \biggl]
	  \nonumber \\
	  &&
	\times  \text{Im} \bigl[ \left< u_{\vec{k} \epsilon  \zeta } \right| P_{\varsigma} \bm{\sigma} P_{\varsigma} \left|  u_{\vec{k} \epsilon'  \zeta' } \right>   
	\left<  u_{\vec{k} \epsilon'  \zeta' } \right| e \vec{E} \cdot   \vec{v}   \left|  u_{\vec{k} \epsilon  \zeta } \right> \bigl], 
 \label{inter}
\end{eqnarray} 
%
where $V$ is the volume of the system, $\left|  u_{\vec{k} \epsilon   \zeta  } \right> $ denotes eigenket of $\mathcal{H}_\text{L}$, $P_\varsigma$ is the projection operator to the sublattice $\varsigma$ ( $P_{\text{A/B}} = ( \text{I}_\tau \pm \tau_z ) /2$ ), $\vec{v} = \frac{1}{\hbar} \partial \mathcal{H}_{\text{L}} / \partial \vec{k}$ is the velocity operator, $f_{\vec{k} \epsilon  \zeta }$ is the Fermi distribution function, and $\Sigma$ is the disorder-induced energy level broadening, which is negligible in clean AFMs.

According to Eq. \eqref{inter}, an occupied state $\vec{k} \epsilon \zeta$ generates contributions to $\delta  \vec{S}_{\varsigma}^{\text{Berry}}$ through virtual interband transition to an unoccupied state $\vec{k} \epsilon' \zeta'$.
Figure \ref{fig:band} shows the two types of interband transition, each of which generates different contributions to $\delta  \vec{S}_{\varsigma}^{\text{Berry}}$. 
We introduce $\delta \vec{S}_{1 \varsigma}^{\text{Berry}}$ and $\delta \vec{S}_{2 \varsigma}^{\text{Berry}}$ to denote the interband contribution with $( \epsilon', \zeta') = ( \epsilon , - \zeta ) $ (green vertical arrows in Fig. \ref{fig:band}) and $( \epsilon', \zeta') = ( - \epsilon , \pm \zeta )$ (red vertical arrows in Fig. \ref{fig:band}), respectively.

First, we evaluate $\delta  \vec{S}_{ \varsigma}^{\text{Berry}}$ for $\vec{m} =0$.
$\delta  \vec{S}_{ 1 \varsigma}^{\text{Berry}} = 0$ since $f_{\vec{k} \epsilon \zeta} -f_{\vec{k} \epsilon,  -\zeta}   =0$.
Whereas $\delta  \vec{S}_{ 2 \varsigma}^{\text{Berry}}$ survives and results in 
\begin{equation}
  \delta  \vec{S}_{ 2 \varsigma}^{\text{Berry}}  = - \frac{c_z}{2 \pi} \frac{ \alpha m_e}{\hbar^2 J} \vec{n} \times ( \hat{z} \times e \vec{E} ), \label{Eq:colspindensity}
\end{equation}
where $c_z$ is a summation factor which is inversely proportional to a lattice constant along $z$ direction \cite{gammaz}.
The resulting SOT $\vec{T}_\varsigma$ becomes
\begin{equation}
    \vec{T}_{\varsigma} 
    = \frac{ J}{\hbar} \vec{M}_\varsigma \times  \delta  \vec{S}_{ 2 \varsigma}^{\text{Berry}}
    = \pm \tau^{\text{Berry}}_\varsigma \vec{n} \times \bigl[ \vec{n} \times ( \hat{z} \times e \vec{E} ) \bigl],
\end{equation}
where $\tau^{\text{Berry}}_\varsigma = - \frac{c_z}{2 \pi} \frac{ \alpha m_e}{\hbar^3}$, and $+(-)$ sign denotes A(B) sublattice.
This agrees with the damping-like SOT reported in Refs. \cite{PhysRevLett.113.157201,PhysRevB.95.014403} for the ideal case $\vec{m} =0$.

\begin{figure}[t!]
\subfigure{\label{fig:LAFMsvm}}
\subfigure{\label{fig:LAFMsvs}}  
\includegraphics[width=8.5cm]{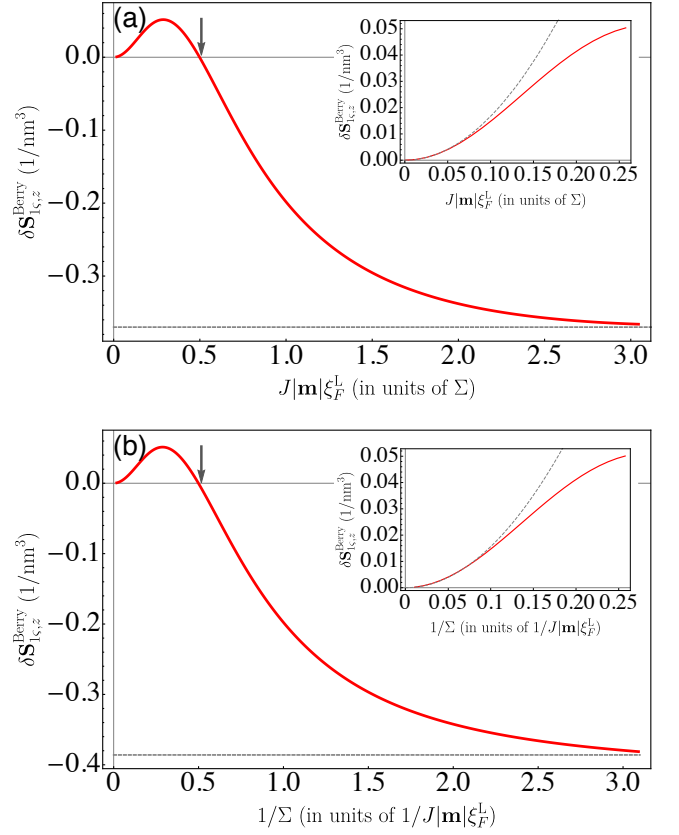}  
\caption{Numerical calculation of the layered AFM for $\delta S^{\text{Berry}}_{1\varsigma,z}$ as function of (a) $J |\vec{m}| \xi_F^{\text{L}}$, and (b) $1 / \Sigma$ being proportional to the momentum scattering time for $m_e^* = 0.013 m_e$, with $m_e$ being mass of electron, and $J= 1 \text{ eV}$, $\alpha = 0.1 \text{ eV}\cdot\text{nm}$, $k_B T =  0.025 \text{ eV}$. We set $\Sigma = 0.01 \text{eV}$ in (a), $J |\vec{m}| = 0.1 \text{ eV}$ in (b). Also, $\mathcal{E}_F = -0.5 \text{ eV}$, $c_z = 1 \text{ nm}^{-1}$, and $e E = 1 \text{ eV nm$^{-1}$}$. Furthermore, we assume that $\gamma(k_z)$ is constant value, and $10^{-2} \text{ eV}$. Insets show that when $J | \vec{m}| \xi^{\text{L}}_F \ll \Sigma$, $\delta S^{\text{Berry}}_{1\varsigma,z}$ is proportional to $( J | \vec{m}| \xi^{\text{L}}_F)^2$ (a), or $(1/\Sigma)^2$ (b).
}
\label{fig:continous}
\end{figure}

Next we evaluate $\delta  \vec{S}_{\varsigma}^{\text{Berry}}$ for small but nonvanishing $\vec{m}$ ($|\vec{m}| \ll 1$).
There appear two interesting limits, which are distinguished from the other by the relative magnitude of $J| \vec{m}| \xi_F^{\text{L}}$ with respect to other energy scales.
Although $\vec{m}$ is small, $J$ is a large energy scale and thus $J| \vec{m}| \xi_F^{\text{L}}$ may or may not be the smallest energy scale of the problem.
In the first limit, we assume $J | \vec{m}| \xi_F^{\text{L}}$ to be much smaller than most energy scales such as $J$, the band width, and the Fermi energy, but larger than the disorder broadening $\Sigma$ and the thermal broadening $k_B T$.
Since $\Sigma$ and $k_B T$ are typically much smaller than $J$ and the Fermi energy, this assumption can be satisfied for small but not too small $|\vec{m}|$.
In this situation, small $\vec{m}$ generates the first-order (in $\vec{m}$) correction to $\delta  \vec{S}_{2 \varsigma}^{\text{Berry}}$, whose effect amounts to replacing $\vec{n}$ in Eq. \eqref{Eq:colspindensity} by $\pm \vec{M}_{\varsigma} = \vec{n} \pm \vec{m}$.
Here $\pm$ sign applies to the A/B sublattices.
This small $\vec{m}$ does not modify $\delta  \vec{S}_{2 \varsigma}^{\text{Berry}}$ in any serious way.
In contrast, small $\vec{m}$ can cause drastic change to $\delta  \vec{S}_{1 \varsigma}^{\text{Berry}}$ in this limit.
When the two-fold degeneracy is lifted by small but nonvanishing $\vec{m}$, $f_{\vec{k} \epsilon \zeta} -f_{\vec{k} \epsilon,  -\zeta}$ becomes finite for small volume of $\vec{k}$.
For those $\vec{k}$, the factor $( E_{\vec{k} \epsilon \zeta} - E_{\vec{k} \epsilon', -\zeta} + i \Sigma)^{-2}$ in summand of Eq. \eqref{inter} diverges as $|\vec{m}|^{-2}$ for small $\Sigma$.
The combined effect of the small $\vec{k}$ volume and the diverging summand generates a contribution which is of \textit{zeroth} order in $\vec{m}$; 
%
\begin{equation}\label{Eq:S1}
 \delta \vec{S}^{\text{Berry}}_{1 \varsigma}  
  =
 - \frac{c_z }{2 \pi} \frac{ \alpha m_e}{ \hbar^2 J  }  \hat{m} \cdot ( \hat{z} \times e \vec{E} ) \hat{m} \times \vec{n},
\end{equation}
where $\hat{m} = \vec{m} / | \vec{m}|$.
%
Together with $\delta  \vec{S}_{2 \varsigma}^{\text{Berry}}$, one obtains the total Berry phase contribution
\begin{eqnarray}\label{Eq:noncolspin}
 \delta \vec{S}^{\text{Berry}}_{\varsigma}  
 &=&
 - \frac{c_z }{2 \pi} \frac{ \alpha m_e}{ \hbar^2 J  }  
 \biggl[
          \hat{m} \cdot ( \hat{z} \times e \vec{E}) \phantom{.} \hat{m} \times \vec{n} 
        + \vec{n} \times ( \hat{z} \times e \vec{E})   
        \nonumber \\
        && \pm  \vec{m} \times ( \hat{z} \times e \vec{E})  
 \biggl] 
 + \mathcal{O}(| \vec{m}|^2).
\end{eqnarray}
Note that even for small $\vec{m}$ with $|\vec{m}| \ll 1$, the first term is comparable in magnitude to the second term reported before \cite{PhysRevLett.113.157201,PhysRevB.95.014403}.
This confirms that the antiunitary symmetry breaking can affect SOT significantly.
A remark is in order.
$\delta \vec{S}^{\text{Berry}}_{1\varsigma}$ in Eq. \eqref{Eq:S1} (or the first term in Eq. \eqref{Eq:noncolspin}) assumes $J | \vec{m}| \xi_F^{\text{L}} \gtrsim \Sigma, k_B T$.
For clean AFM with $\Sigma = 0.01$ eV and at room temperature with $k_B T = 0.026$ eV, this inequality is satisfied for $J |\vec{m}| \xi_F^{\text{L}} \sim 0.01$.

In the opposite limit, where $J | \vec{m}| \xi_F^{\text{L}} \ll \text{max}[ \Sigma, k_B T]$, Eq. \eqref{Eq:S1} acquires the extra factor $( J \vec{m} \xi_F^{\text{L}} )^2 / \text{max}^2[ \Sigma, k_B T]$ and thus becomes proportional to $\vec{m}^2$. 
For general ratio between $J | \vec{m}| \xi_F^{\text{L}} $ and $\text{max}[ \Sigma, k_B T]$, it is difficult to carry out analytic calculation.
Figure \ref{fig:continous} shows the numerical calculation results of $\delta \vec{S}_{1 \varsigma}^{\text{Berry}} \cdot \hat{z}$ as a function of $J  |\vec{m}| \xi_F^{\text{L}}$ (Fig. \ref{fig:LAFMsvm}) and $1/ \Sigma$ (Fig. \ref{fig:LAFMsvs}).
Here, $\vec{n} = |\vec{n}| \hat{x}$, $\vec{m} = | \vec{m}| \hat{y}$, and $|\vec{n}|^2 + | \vec{m}|^2 = 1$.
The horizontal dashed lines in Figs. \ref{fig:LAFMsvm} and \ref{fig:LAFMsvs} denote the asymptotic behavior in the large $J|\vec{m}| \xi_F^{\text{L}} / \Sigma$ limit given in Eq. \eqref{Eq:S1} and the quadratically growing dahsed lines in the insets of Figs. \ref{fig:LAFMsvm} and \ref{fig:LAFMsvs} the limiting behavior in the small $J | \vec{m}| \xi_F^{\text{L}} / \Sigma$ limit.
Interestingly, there occur sign changes both in Figs. \ref{fig:LAFMsvm} and \ref{fig:LAFMsvs} around $J | \vec{m}| \xi_F^{\text{L}} / \Sigma \sim 1/2$.
We remark that $\xi_F^{\text{L}}$ is closely related with $\gamma(k_z)$, which is inter-sublattice hopping parameter.
If $\gamma(k_z)$ is very weak, the gray arrows ($\delta \vec{S}^{\text{Berry}}_{1\varsigma} = 0$ point) in Fig. \ref{fig:continous} are shifted to the right and $|\vec{m}|$ should be larger in order to realize the first limit.

\begin{figure}[t!]
\subfigure{\label{fig:oldDLT}}  
\subfigure{\label{fig:newDLT}}
 
\includegraphics[width=8.5cm]{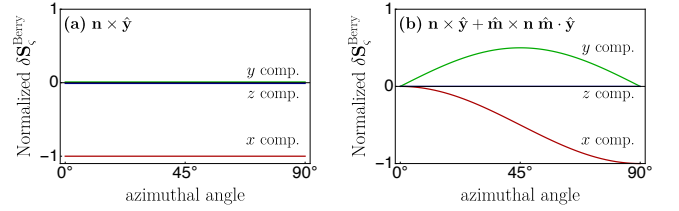}  
\caption{ 
(a), (b) The normalized $\delta \vec{S}^{\text{Berry}}_{\varsigma }$ (first two terms in Eq. \eqref{Eq:noncolspin}) as a function of an azimuthal angle $\varphi$. Each solid line represents $x$, $y$, $z$ components of the normalized $\delta \vec{S}^{\text{Berry}}_{\varsigma }$. Here, $\vec{n} = \hat{z}$, $\hat{m} = (-\sin \varphi, \cos \varphi, 0)$, and $\vec{E} \parallel \hat{x}$.
}
\label{fig:vector}
\end{figure}

%
When Eq. \eqref{Eq:S1} is valid, Eq. \eqref{Eq:noncolspin} implies significant change in the angular dependence of $\delta \vec{S}^{\text{Berry}}_{\varsigma}$.
Figure \ref{fig:newDLT} shows the angular dependence of $\delta \vec{S}^{\text{Berry}}_{\varsigma}$ as a function of the azimuthal angle $\varphi$ of $\vec{m}$, when $\vec{m} = ( - \sin \varphi, \cos \varphi, 0 )$ lies in the xy plane, $\hat{n} = \hat{z}$, and $\vec{E} \parallel \hat{x}$.
For simplicity, only the first two terms in Eq. \eqref{Eq:noncolspin} are taken into account in Fig. \ref{fig:newDLT}.
Note that $\delta \vec{S}^{\text{Berry}}_{\varsigma}$ varies significantly with $\varphi$, which is in clear contrast to the result in the perfect collinear case (Fig. \ref{fig:oldDLT}) with the antiunitary symmetry.

\subsection{\label{subsec:BAFM}Bipartite AFM}

The layered AFM is not the only system where the antiunitary symmetry breaking can affect SOT.
To demonstrate this point, we consider the bipartite AFM (Fig. \ref{fig:BAFM}) as the second example.
The interface between AFM and heavy metal belongs to this structure.
It has been demonstrated \cite{PhysRevB.95.014403} that this system generates SOT when the inversion symmetry P is broken.
Note that both sublattices A and B are subject to the same sign of the inversion symmetry breaking unlike the layered AFM.
Another important difference from the layered AFM is the relevant transformations for the bipartite AFM.
They are T$_{a \hat{x}}$, T$_{a \hat{y}}$, and T, where T$_{a \hat{x}}$, T$_{a \hat{y}}$ denote the translations by $a \hat{x}$ and $a \hat{y}$, respectively, and $a$ is the lattice spacing.
None of these transformation is the symmetry of the system.
However their combinations may be symmetries.
T$_{a \hat{x}}$T$_{a \hat{y}}$ is the symmetry of the system regardless of $\vec{m}$, and T$_{a \hat{x}}$T, T$_{a \hat{y}}$T are the symmetries of the system only for $\vec{m} = 0$.
Of particular importance for SOT are the latter two symmetries, which are antiunitary and broken when $\vec{m}$ becomes nonzero.

\begin{figure}[t!] 
\includegraphics[width=8.5cm]{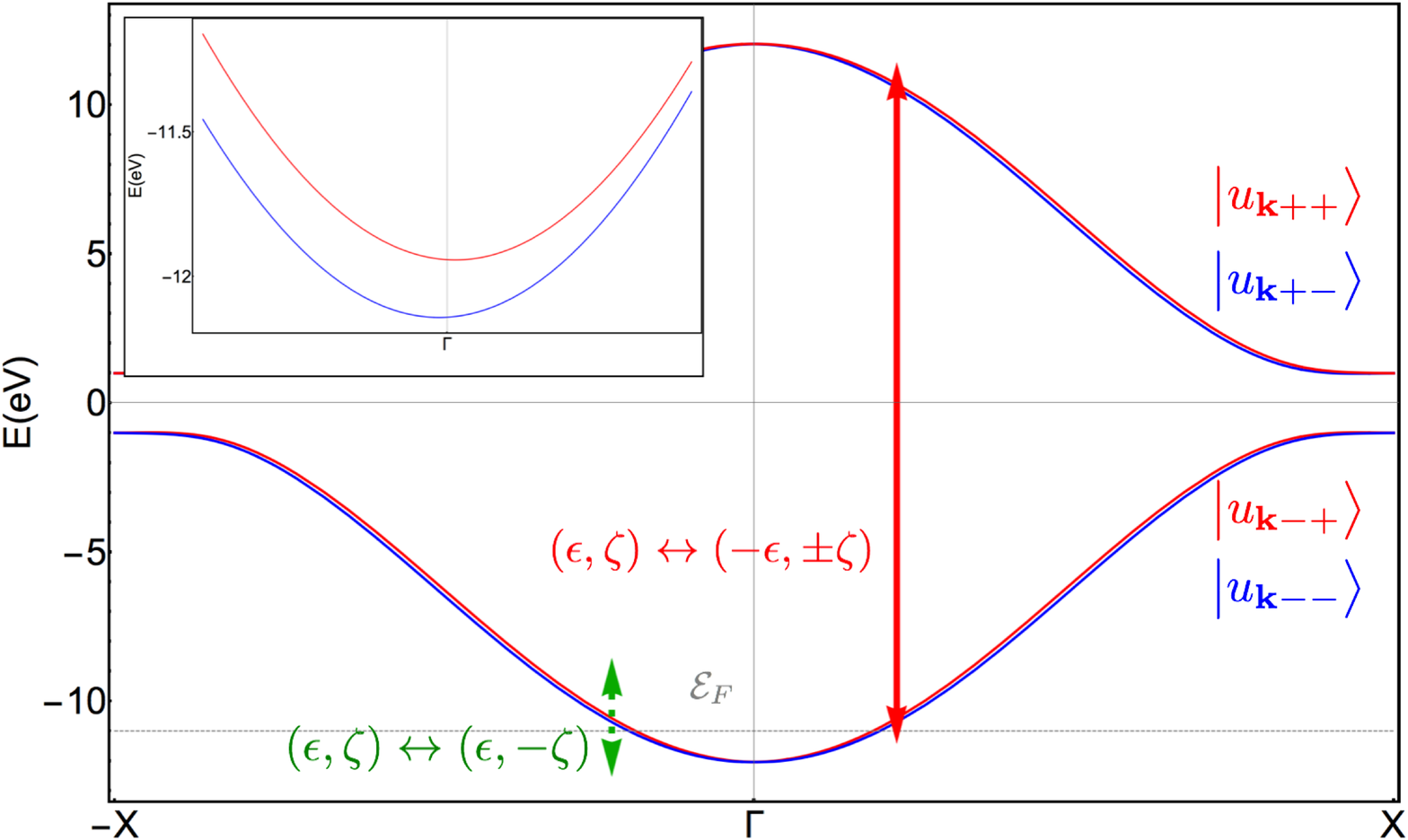} 
\caption{The band structures for the bipartite AFM with $\vec{m} = 0$ (see Appendix). Each arrow denotes $\delta \vec{S}^{\text{Berry}}_{1 \varsigma}$ (green arrow), or $\delta \vec{S}^{\text{Berry}}_{2 \varsigma}$ (red arrow). The dashed green arrow does not contribute to non-equilibrium spin density, i.e., $\delta \vec{S}^{\text{Berry}}_{1 \varsigma} = 0$. Inset shows schemetically two lower bands of the band structure for bipartite AFM with nonvanishing $\vec{m}$. Note that the lower and upper bands are origin symmetric, and the degeneracy even occurs at X point.}
\label{fig:BAFMband}
\end{figure}

In order to demonstrate effects of the nonzero $\vec{m}$, we consider the following four-band Hamiltonian \cite{PhysRevLett.113.157201} for two-dimensional bipartite AFM,
\begin{equation} \label{eq:BAFMH}
 \mathcal{H}_{\text{B}} = 
 \begin{pmatrix}
  J \bm{\sigma} \cdot  \vec{M}_{\text{A}}  & \gamma_{\vec{k}_\parallel} \\
  ( \gamma_{\vec{k}_\parallel} )^* & J  \bm{\sigma} \cdot \vec{M}_{\text{B}} 
 \end{pmatrix}_\tau
 +
 \frac{\alpha}{a} \bm{\sigma} \cdot ( \vec{q} \times \hat{z})  \tau_x,
\end{equation}
where $\gamma_{\vec{k}_\parallel} = - 2 t ( \cos k_x a + \cos k_y a )$ is a hopping energy, and $\vec{q} = ( \sin k_x a, \sin k_y a, 0)$.
Figure \ref{fig:BAFMband} shows the band structure for bipartite AFM with $\vec{m} =0$.
Note that there are two-fold degeneracies at the high symmetry points $\Gamma$, X (and also M, which is not shown) due to the antiunitary symmetries T$_{a \hat{x}}$T, T$_{a \hat{y}}$T.
These symmetries, however, do not impose the two-fold degeneracy for other $\vec{k}$ points, which is in clear contrast to the two-fold degeneracy for all $\vec{k}$ points imposed by the PT symmetry for the layered AFM.
This difference between PT and T$_{a \hat{x}}$T (or T$_{a \hat{y}}$T) arises from the fact that $\vec{k}$ leaves invariant under PT but transforms to $- \vec{k}$ under T$_{a \hat{x}}$T (or T$_{a \hat{y}}$T).
When $\vec{m}$ is nonzero, on the other hand, the antiunitary symmetries are broken and the two-fold degeneracies at high symmetry $\vec{k}$ points are lifted.
The inset in Fig. \ref{fig:BAFMband} shows the degeneracy-lifting at the $\Gamma$ point.

Note that the band structures of the bipartite AFM before and after the antiunitary symmetry breaking by $\vec{m}$ resemble those of the two-dimensional non-magnetic Rashba system before \cite{Aronov:1989tt,*Edelstein:1990ke} and after \cite{kurebayashi2014antidamping,*PhysRevB.91.144401} the time-reversal symmetry breaking by the emergence of FM order.
Then just like the time-reversal symmetry breaking can drastically modify the current-induced spin polarization in the Rashba system, it is reasonable to expect that the antiunitary symmetry breaking of the bipartite AFM by $\vec{m}$ may modify the non-equilibrium spin density $\delta \vec{S}^{\text{Berry}}_{ \varsigma}$ generated by the current significantly.

Similarly to the situation for the layered AFM, $\delta \vec{S}^{\text{Berry}}_{ \varsigma}$ has two contributions, $\delta \vec{S}^{\text{Berry}}_{1 \varsigma}$ and $\delta \vec{S}^{\text{Berry}}_{2 \varsigma}$, which arise from the interband transitions marked by the green and red arrows in Fig. \ref{fig:BAFMband}, respectively.
For $\vec{m} = 0$, $\delta \vec{S}^{\text{Berry}}_{1 \varsigma}$ vanishes due to the antiunitary symmetries T$_{a \hat{x}}$T and T$_{a \hat{y}}$T (See Appendix) whereas 
\begin{equation} \label{Eq:coBAFMspin}
 \delta \vec{S}^{\text{Berry}}_{2 \varsigma} = \mp S_\vec{n} \hat{n} \times ( \hat{z} \times e \vec{E}),
\end{equation}
where - (+) sign applies to $\varsigma =$ A (B) sublattice.
Equation \eqref{Eq:coBAFMspin} has been obtained before \cite{PhysRevB.95.014403}.
$S_\vec{n}$ is generally propertional to $\alpha J / \mathcal{E}_F^2$, where the $1 / \mathcal{E}_F^2$ dependence comes from the fact that the energy level spacing for the red-arrow-marked interband transitions (Fig. \ref{fig:BAFMband}) is $2 \mathcal{E}_F$. 
The precise value of $S_\vec{n}$ depends on various details.
When $\mathcal{E}_F$ is near the bottom of the lower energy bands as in Fig. \ref{fig:BAFMband}, $S_\vec{n} = ( m^* / 4 \pi \hbar^2 ) ( \alpha J / \mathcal{E}_F^2)$, where $m^*$ is the effective mass at the $\Gamma$ point.

\begin{figure}[t!] 
\includegraphics[width=8.5cm]{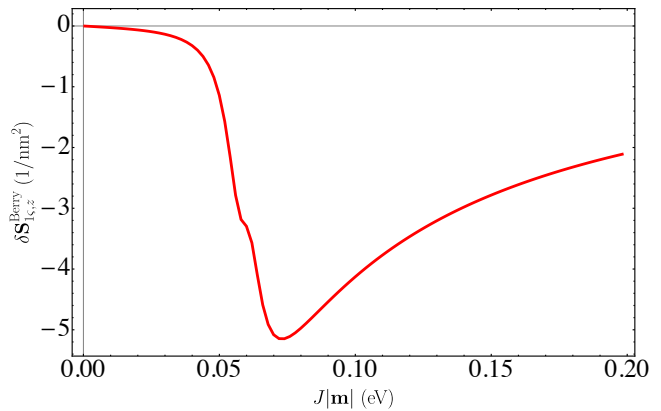}  
\caption{ 
Numerical calculation of the bipartite AFM for $\delta S^{\text{Berry}}_{1\varsigma,z} $ as function of $J |\vec{m}|$ for $t = 3 \text{ eV}$, $J= 1 \text{ eV}$, $\alpha = 0.1 \text{ eV}\cdot\text{nm}$, $k_B T =  0.026 \text{ eV}$. We set $\Sigma = 0.01 \text{ eV}$, $eE = 1 \text{ eV nm$^{-1}$}$, and $\mathcal{E}_F = -11 \text{ eV}$ (the corresponding Fermi momentum is $k_F \simeq 0.6 \text{ nm}^{-1}$). Here, $\vec{n}$ is parallel to $\hat{y}$ direction, and $\vec{m}$ is $-\hat{x}$ direction.
}
\label{fig:BAFMsvm}
\end{figure}

For small but nonvanishing $\vec{m}$, $\delta \vec{S}^{\text{Berry}}_{2 \varsigma}$ acquires a correction of order $\mathcal{O}(|\vec{m}|)$, which stays small for $|\vec{m}| \ll 1$.
We thus ignore this correction.
On the other hand, $\delta \vec{S}^{\text{Berry}}_{1 \varsigma}$ may acquire a more significant correction.
As remarked above $\delta \vec{S}^{\text{Berry}}_{1 \varsigma}$ arises from the green-arrow-marked interband transition in Fig. \ref{fig:BAFMband} and the antiunitary symmetry constrains (Appendix) that prevent $\delta \vec{S}^{\text{Berry}}_{1 \varsigma}$ do not work anymore since the symmetries are broken once $\vec{m}$ is not zero.
Generally, Berry phase effects are known to be larger when they arise from interband transitions of smaller energy spacing.
In that respect, it is not surprising to have sizable $\delta \vec{S}^{\text{Berry}}_{1 \varsigma}$ since the green-arrow-marked interband transition involves small energy spacing; the spacing is $\bigl( 2 | \gamma_{\vec{k}_{\parallel}} | / \sqrt{ J^2 + | \gamma_{\vec{k}_\parallel}|^2} \bigl) J | \vec{m}|$ for $J | \vec{m}| \gg \alpha k_F$ and $2 \alpha k_F \Delta_\vec{k}$ for $J | \vec{m} | \ll \alpha k_F$, where $k_F$ is the Fermi wavevector, $\Delta_\vec{k}$ is the anisotropy parameter (See Appendix).
In both cases, it is much smaller than the level spacing for the red-arrow-marked interband transition (Fig. \ref{fig:BAFMband}).
We first calculate $\delta \vec{S}^{\text{Berry}}_{1  \varsigma}$ in the two opposite limits.
In the first limit, where both $J | \vec{m}|$ and $\alpha k_F$ are sufficiently larger than $\text{max}( \Sigma, k_B T)$, the antiunitary symmetry constraints become completely obsolete and the green-arrow-marked interband transition produces a sizable contribution, whose magnitude depends rather sensitively on the direction of $J \vec{m}$ and its relative magnitude with respect to $\alpha k_F$.
For concreteness, we assume both $\vec{m}$ and $\vec{n}$ to be within the $xy$ plane.
Then for $J | \vec{m}| \gg \alpha k_F$, we obtain
\begin{equation}
  \delta \vec{S}^{\text{Berry}}_{1  \varsigma} = - S_{\vec{m}} \hat{m} \times ( \hat{z} \times e \vec{E} ), \label{eq:BAFMnonS}
\end{equation}
where $S_\vec{m}$ is approximately $S_\vec{n}  (\mathcal{E}_F^2 / J^2 | \vec{m}| )$.
Note that $ | \delta \vec{S}^{\text{Berry}}_{1  \varsigma}| /   |\delta \vec{S}^{\text{Berry}}_{2  \varsigma}| \sim \mathcal{E}_F^2 /   J^2 | \vec{m}|  $ is \textit{inversely} proportional to $\vec{m}$, that is, for small $\vec{m}$, the SOT correction due to the antiunitary symmetry breaking may be even larger than the previously reported SOT \cite{PhysRevB.95.014403} in the ideal limit with the antiunitary symmetry.
The situation is less drastic for $\alpha k_F \gg J | \vec{m}|$, when a simple order counting results in $| \delta \vec{S}^{\text{Berry}}_{1  \varsigma}| / |\delta \vec{S}^{\text{Berry}}_{2  \varsigma}|  \sim   | \vec{m}|  (\mathcal{E}_F  /  \alpha k_F)^2$.
This ratio is proportional to $|\vec{m}|$ but its proportionality constant $\sim ( \mathcal{E}_F / \alpha k_F )^2$ can easily go over $100$ or more.
Thus, even in this case, the SOT correction due to the antisymmetry breaking can be significant.

In the opposite limit, where either $J| \vec{m}|$ or $\alpha k_F$ is sufficiently smaller than $\text{max}(\Sigma, k_B T)$, the antiunitary symmetry constraints remain effective and the Berry phase effect from the green-arrow-marked interband transition is suppressed.
Figure \ref{fig:BAFMsvm} shows the numerical calculation results of $\delta \vec{S}^{\text{Berry}} \cdot \hat{z}$ as function of $J | \vec{m}|$.
In this calculation, it is assumed that $\alpha k_F$ is about factor 10 larger than $\Sigma$ and $k_B T$.
Note that starting from $J |\vec{m}| \sim 0.07$, for which $| \vec{m}| \simeq 0.07$, $\delta \vec{S}^{\text{Berry}}_{1  \varsigma}$ in Fig. \ref{fig:BAFMsvm} follows the $1 / | \vec{m}|$ dependence predicted by Eq. \eqref{eq:BAFMnonS}.
For $J | \vec{m}| \lesssim 0.04$, on the other hand, $\delta \vec{S}^{\text{Berry}}_{1  \varsigma}$ is strongly suppressed as expected.
Unlike the sign change in the layered AFM (Fig. \ref{fig:continous}), the sign change does not occur in the bipartite AFM (Fig. \ref{fig:BAFMsvm}).
%


\section{\label{sec:Discussion}Discussion}

As illustrated in Figs. \ref{fig:LAFMsvm} and \ref{fig:BAFMsvm}, $\delta \vec{S}^{\text{Berry}}_{1 \zeta} \cdot \hat{z}$ in the layered AFM and the bipartite AFM depends on $\vec{m}$ in different way; 
When $\vec{m}$ is sufficiently large to overcome the level broadening (but smaller than $\vec{n}$), $\delta \vec{S}^{\text{Berry}}_{1 \zeta} \cdot \hat{z}$ in the layered AFM saturates to a value that is independent of $|\vec{m}|$ (Fig. \ref{fig:LAFMsvm}) but $\delta \vec{S}^{\text{Berry}}_{1 \zeta} \cdot \hat{z}$ in the bipartite AFM is inversely proportional to $|\vec{m}|$.
This difference originates from the symmetry difference between the two types of AFMs.
In the layered AFM, the energy dispersion is two-fold degenerate for $\vec{m} = 0$ due to the PT symmetry and this degeneracy is lifted by $\vec{m}$ (Fig. \ref{fig:band}).
In the bipartite AFM, on the other hand, the energy dispersion is {\it not} two-fold degenerate even for $\vec{m} =0$ because the T$_{a \hat{x}}$T or T$_{a \hat{y}}$T symmetries do not guarantee the degeneracy for general $\vec{k}$.
The symmetries guarantee the degeneracy only at high symmetry points such as $\Gamma$ and X.
When $\vec{m}$ becomes finite, the symmetries are broken and even this degeneracy is lifted.
The resulting energy dispersion resembles that for a two-dimensional FM Rashba model (Fig. \ref{fig:BAFMband}).
Thus it is reasonable to expect that for nonzero $\vec{m}$, $\delta \vec{S}^{\text{Berry}}_{1 \varsigma} \cdot \hat{z}$ for the bipartite AFM is similar to that the two-dimensional FM Rashba model which is already well known \cite{kurebayashi2014antidamping,*PhysRevB.91.144401, PhysRevB.91.134402}.
In this analogy with the two-dimensional FM Rashba model \cite{kurebayashi2014antidamping,*PhysRevB.91.144401, PhysRevB.91.134402}, $J | \vec{m}|$ in the bipartite AFM corresponds to $J M_s$ (which is commonly abbreviated to $J$ since $M_s$ is essentially a constant in FM systems) in the FM Rashba model.
Then, considering that $\delta \vec{S}$ generated by the Berry phase in the FM Rashba model is inversely proportional to $J M_s$ \cite{kurebayashi2014antidamping,*PhysRevB.91.144401, PhysRevB.91.134402}, $\delta \vec{S}^{\text{Berry}}_{1 \varsigma}$ being inversely proportional to $| \vec{m}|$ in the bipartite AFM is natural.
For the layered AFM, such analogy with the two-dimensional FM Rashba system is not possible since the energy dispersion has very different structure.

We address the question of whether the crucial assumption $J | \vec{m}| \xi_F^{\text{n}} > \Sigma,\phantom{.} k_B T$ can be satisfied in experiments, where $\text{n}$ denotes either L (layered AFM) or B (bipartite AFM).
Here, $\xi_F^{\text{n}}$ is a dimensionless parameter.
We note that $\xi_F^{\text{L}}$ is proportional to $\gamma(k_z)$ in the layered AFM, since we assume $\gamma(k_z) < J$.
In the bipartite AFM, when $J | \vec{m}| \gg \alpha k_F$, one obtains $\xi_F^{\text{B}} \sim | \gamma_{\vec{k}_{\parallel}}| / \sqrt{ J^2 + \gamma^2_{\vec{k}_{\parallel}}}$.
For $\Sigma \sim 0.01 \text{ eV}$ \cite{PhysRevB.91.134402}, $J = 1 \text{ eV}$, $\xi_F^{\text{L}} \sim 0.1$ ($\xi_F^{\text{B}} \sim 1$), and at room temperature $k_B T \simeq 0.026 \text{ eV}$, the assumption requires $| \vec{m}| \gtrsim 0.1$ ($0.01$), which amounts to the spin canting of $\gtrsim 10 \%$ $(1 \%)$.
We remark that $\xi_F^{\text{n}}$ depends on $J$, and the inter-sublattice hopping, such as $\gamma(k_z)$ and $\gamma_{\vec{k}_\parallel}$.
Even for such AFMs, $\vec{m}$ becomes finite during magnetization dynamics and can be estimated as $\vec{m} \sim \hbar / J_H | \partial_t \vec{n}|$, where $J_H$ is exchange parameter between local mangnetic moments \cite{PhysRevB.83.054428}.
Thus the assumption amounts to  $ \hbar  J / J_H | \partial_t \vec{n}| > \Sigma, \phantom{.} k_B T$.
For example, in Mn$_2$Au \cite{APL.93.162503}, $J_H \sim 0.04 \text{ eV}$. 
For the material parameters specified above, the assumption is satisfied for fast magnetization dynamics with characteristic frequency $\gtrsim 1  \text{ THz}$.
Considering that the characteristic magnetization dynamics in AFMs is estimated to be in THz scale, this estimation indicates that the antiunitary symmetry breaking effect on SOT can be indeed relevant for fast AFM magnetization dynamics.

We consider two types of AFM magnetization dynamics. 
One is the AFM magnetization switching dynamics. 
A recent experiment \cite{Wadleyaab1031} reported the electrical switching of the AFM magnetization through SOT. 
The time scale of the dynamics is quite slow, however. 
In Ref. \cite{Wadleyaab1031}, the current pulse length for the AFM magnetization switching should be longer than 1 ms, which implies that the characteristic frequency of the AFM magnetization switching is of the order of 1 kHz at best, which is far below the required frequency of 1 THz mentioned above. 
We thus conclude that the effect of the noncollinearity is irrelevant for the slow AFM magnetization switching realized in Ref. \cite{Wadleyaab1031}. 
The other type of the AFM magnetization dynamics we consider is the AFM domain wall motion. Recent theoretical analyses \cite{PhysRevLett.117.017202,*PhysRevLett.117.087203} predict AFM domain walls to move at high speeds of a few to a few tens of km/s.
For the domain wall width of a few nm \cite{PhysRevLett.117.017202,*PhysRevLett.117.087203}, the corresponding frequency scale of the magnetization dynamics at the center of AFM domain walls ranges $1 \sim 10 \text{ THz}$.
Thus for such a high speed AFM domain wall motion, we expect that the noncollinearity effect may become relevant.

Still another way to test the noncollinearity effect is to measure SOT for AFMs whose equilibrium spin configurations are noncollinear. 
Such equilibrium noncollinearity may be weakly induced by the Dzyaloshinskii-Moriya interaction \cite{PhysRev.120.91,PhysRevB.71.184434,dmitrienko2014measuring}, which is often present in AFMs. 
Another way to induce the equilibrium noncollinearity is to apply an external magnetic field. This is probably the simplest way at least in principle. 
But for conventional AFMs with $J_H \sim 0.04 \text{ eV}$, the required field strength is of the order of $100$ T or more, which may be too high. 
Thus to induce the equilibrium noncollinearity by an external field, a synthetic AFM \cite{PhysRevB.95.104435} will be a more suitable system since the corresponding $J_H$ will be much smaller.

Lastly we remark on a technical limitation of our calculation. 
Our calculation takes the disorder effect into account only through the self-energy correction via the energy level broadening. 
However the disorder may affect the calculation result also through the vertex correction, which is completely ignored in this paper. 
For certain spin-related phenomena such as the spin Hall effect in the two-dimensional Rashba model \cite{PhysRevLett.92.126603}, it has been reported that the vertex correction completely cancels \cite{PhysRevB.70.041303,PhysRevLett.93.226602} the spin Hall effect predicted by the calculation \cite{PhysRevLett.92.126603} that ignores the vertex correction. 
Thus in certain cases, the neglect of the vertex correction can be dangerous. 
However this appears to be rather exceptional. 
For modified Rashba models for which the energy dispersion deviates \cite{PhysRevB.73.195307} from the quadratic $k$ dependence or the spin-momentum coupling deviates \cite{PhysRevB.71.121308} from the $k$-linear dependence in the ideal Rashba model, the complete cancellation by the vertex correction does not occur. 
Moreover for $p$-type semiconductors, whose Hamiltonian deviates significantly from the ideal Rashba model, the vertex correction itself is found to be absent \cite{PhysRevB.69.241202}. 
It has been argued \cite{sinova2006spin} that a complete cancellation by the vertex correction is limited to the two-dimensional Rashba model and does not occur in general. 
Based on previous studies on the vertex correction, we think it is unlikely for the vertex correction to qualitatively modify the noncollinearity effect predicted in this paper although quantitative correction is likely. 
To verify this expectation, an explicit calculation of the vertex correction is necessary, which however goes beyond the scope of this paper.

\section{\label{sec:Summary}Summary}
In summary, for two types of AFMs we have demonstrated that small deviation from perfectly collinear spin configurations can significantly modify properties of SOT.
This result originates from the fact that the noncollinearity breaks antiunitary symmetries present when spin configurations are collinear.
Due to this connection with antiunitary symmetries we expect this result to hold for more general situations including other types of AFMs with broken antiunitary symmetries.
For instance, antiunitary symmetries are broken in equilibrium if AFMs have weak ferrimagnetic character or in AFMs with the broken inversion symmetry where the Dzyaloshinskii-Moriya interaction results in canting and breaks antiunitary symmetries.
A similar situation may appear at a bilayer that consists of a AFM and a heavy metal or of an AFM and a FM layer. 
We thus expect this effect to be generic in various classes of AFMs where antiunitary symmetries are broken either dynamically or in equilibrium.
This result will be relevant for the realization of THz scale AFM devices.

\begin{acknowledgments}
We acknowledge useful discussion with Kyung-Jin Lee.
This work was the support by the National Research Foundation of Korea (NRF) (Grant No. 2011-0030046) and the MOTIE (Grant No. 10044723).
\end{acknowledgments}


\revappendix*
\section{}\label[appendix]{app:sym}

Considering Eq.\eqref{eq:BAFMH} with $\vec{m} = 0$, the the corresponding eigenvalues and eigenstates are as follows:
\begin{equation}
 \mathcal{H}_{\text{B}} \bigl| u_{\vec{k}\epsilon \zeta } \bigl> 
 =  
 E_{\vec{k}\epsilon \zeta }  \bigl| u_{\vec{k}\epsilon \zeta } \bigl>,  
\end{equation}  
where $E_{\vec{k} \epsilon \zeta }  =  \epsilon  \sqrt{ \mathcal{E}^2_0(\vec{k}) + \epsilon \zeta  2 \alpha  \mathcal{E}_0 (\vec{k})  \Delta_\vec{k} + \alpha^2 q^2 }$, $\mathcal{E}_0(\vec{k}) = | E_{\vec{k} \epsilon \zeta } (\alpha  =0)|$, and $\Delta^2_{\vec{k}} =  q^2 - J^2/\mathcal{E}_0^2  [\hat{n} \cdot ( \vec{q} \times \hat{z} ) ]^2  $ \cite{PhysRevB.95.014403}.

For the convenience of symmetry analysis, the Bloch wave function may be written as
\begin{equation}
 \bigl| u_{\vec{k} \epsilon \zeta} \bigl>  = e^{ i \vec{k} \cdot \vec{r} } \bigl| u_{\text{R} \epsilon \zeta } \bigl> + e^{ - i   \vec{k} \cdot \vec{r} } \bigl| u_{\text{L} \epsilon \zeta }  \bigl>, \label{dl+} 
\end{equation}
where $\vec{k}>0$. 
The eigenstates are related by operations: 
\begin{equation}
 \text{T}_\vec{k} \text{T} 
 \bigl| u_{\text{R} \epsilon \zeta } \bigl> 
  =  
 e^{-i \vec{k} \cdot \vec{a} } 
 \bigl| u_{\text{L} \epsilon \zeta } \bigl>. \label{eq:tkt} 
\end{equation}

Let us consider the imaginary part of Eq. \eqref{inter} for interband process in Fig. \ref{fig:BAFMband}. 
We assume that the Fermi energy is positioned at the bottom of the lower energy bands.
In the interband transition between the two lower bands, the spin and velocity parts are
\begin{eqnarray}
&&\bigl< u_{\text{L}-+} \bigl| \bm{\sigma}_{\text{A/B}} \bigl| u_{\text{L}--} \bigl> 
 = 
-  e^{ i \vec{k} \cdot 2 \vec{a}}  \bigl< u_{\text{R}-+} \bigl|  \bm{\sigma}_{\text{B/A}}    \bigl|u_{\text{R}--}  \bigl>^*, \nonumber 
\\
\\
&& \bigl< u_{\text{L}--} \bigl| \hat{v} \bigl| u_{\text{L}-+} \bigl>
 = 
- e^{  - i \vec{k} \cdot 2 \vec{a}} \bigl<    u_{\text{R}--} \bigl| \hat{v} \bigl| u_{\text{R}-+} \bigl>^*,
\end{eqnarray}
where $\bm{\sigma}_{\text{A/B}} = P_{\text{A/B}}\bm{\sigma} P_{\text{A/B}}$, and 
\begin{align}
( \text{T}_\vec{k} \text{T} )^{-1} \bm{\sigma}_{A/B} ( \text{T}_\vec{k} \text{T} )
= - \bm{\sigma}_{B/A}, \phantom{..} 
( \text{T}_\vec{k} \text{T} )^{-1}  \hat{v} ( \text{T}_\vec{k} \text{T} )
=
- \hat{v}. 
\end{align}
So, one obtains
\begin{eqnarray}\label{RL}
 && \text{Im} \bigl[ \bigl< u_{\text{L}-+} \bigl|   \bm{\sigma}_{\text{A/B}} \bigl| u_{\text{L}--} \bigl> \bigl< u_{\text{L}--} \bigl| e \vec{E} \cdot   \vec{v}   \bigl| u_{\text{L}-+} \bigl> \bigl] 
 \nonumber \\
 && \phantom{.}
  = 
  -  \text{Im} \bigl[ \bigl< u_{\text{R}-+} \bigl|   \bm{\sigma}_{\text{B/A}} \bigl| u_{\text{R}--} \bigl> \bigl< u_{\text{R}--} \bigl| e \vec{E} \cdot   \vec{v}   \bigl| u_{\text{R}-+} \bigl> \bigl].
  \nonumber \\
\end{eqnarray}
Furthermore, $\text{P} \bigl| u_{L \epsilon+}^i \bigl>$, $\bigl| u_{R \epsilon-}^i \bigl>$ have same eigenvalue, because $\text{P}^{-1} \mathcal{H}_{\text{B}}(\vec{k}) \text{P} = \mathcal{H}_{\text{B}}(- \vec{k})$. Thus,
\begin{eqnarray}
   &&  \bigl<  u_{\text{L}-+}   \bigl| \bm{\sigma}_{\text{A/B}}  \bigl| u_{\text{L}--}  \bigl>
   \nonumber \\
    && \phantom{asdf} =  
     \bigl<  u_{\text{L}-+}   \bigl| \bm{\sigma}_{\text{A/B}} \text{T}_\vec{k} \text{T} \text{P} \bigl| u_{\text{L}-+}  \bigl>
     \\
    && \phantom{asdf} =   
     \bigl<  u_{\text{L}-+}   \bigl|  \text{T}_\vec{k} \bm{\sigma}_{\text{B/A}}   \text{T} \text{P} \bigl| u_{\text{L}-+}  \bigl>
      \\
    && \phantom{asdf} =  
     \bigl<\bm{\sigma}_{\text{B/A}}  \text{T}_\vec{k}^\dag  u_{\text{L}-+}  \bigl|   \text{T} \text{P}   u_{\text{L}-+}  \bigl>
     \\
    && \phantom{asdf} =  
     \bigl< \text{T} \text{P}   u_{\text{L}-+} \bigl| T^{-1} \bm{\sigma}_{\text{B/A}}  \text{T}_\vec{k}^\dag  u_{\text{L}-+}   \bigl>
     \\
    && \phantom{asdf} =  
     \bigl<  u_{\text{L}-+}  \bigl|   \bm{\sigma}_{\text{B/A}}  \text{T}_\vec{k} \text{T} \text{P} \bigl| u_{\text{L}-+}  \bigl>, \label{abba}
\end{eqnarray}
where $\text{P} \bm{\sigma}_{\text{A/B}} = \bm{\sigma}_{\text{A/B}} \text{P} $, $\text{T}_\vec{k} \bm{\sigma}_{\text{A/B}} \text{T}_{\vec{k}}^{-1} = \bm{\sigma}_{\text{B/A}}$, PT$^{-1}_{\vec{k}} = \text{T}_\vec{k}$P.
Consequently, Eqs. \eqref{RL}, \eqref{abba} indicate that the Berry phase contribution between $\bigl| u_{\text{L}-+} \bigl>$ and $\bigl| u_{\text{L}--} \bigl>$ are exactly cancelled by the Berry phase contribution between $\bigl| u_{\text{R}-+} \bigl>$ and $\bigl| u_{\text{R}--} \bigl>$.

\nocite{*} 
\bibliography{manuscript.bbl}

\begin{thebibliography}{39}
\bibitem [{gam()}]{gammaz}%
  \BibitemOpen
  \href@noop {} {}\bibinfo {note} {We assume that $\gamma(k_z)$ is small
  parameter. Then, $\delta \vec{S}_{2 \varsigma} = - J \alpha m_e / ( 2 \pi
  \hbar^2) \vec{n} \times ( \hat{z} \times e \vec{E}) \bigl[ \int (dk_z/2 \pi)
  ( J^2 + \gamma^2(k_z) )^{-1} \bigl] + \mathcal{O}(\gamma^2(k_z) )$. Here the
  square bracket can be written as $(1 / J^2) \int (dk_z/ 2 \pi) +
  \mathcal{O}(\gamma^2(k_z)) \sim c_z / J^2$. We remark that the angular
  dependence does not change, though $\gamma(k_z)$ is quite large.}\BibitemShut
  {Stop}%
\end{thebibliography}

\end{document}